\documentclass[sigconf]{acmart}

\usepackage{xac}
\usepackage{wrapfig}
\usepackage{caption}
\usepackage{subcaption}

\AtBeginDocument{%
  \providecommand\BibTeX{{%
    \normalfont B\kern-0.5em{\scshape i\kern-0.25em b}\kern-0.8em\TeX}}}

\setcopyright{acmcopyright}
\copyrightyear{2021}
\acmYear{2021}
\setcopyright{acmcopyright}
\acmConference[IUI '21]{26th International Conference on Intelligent User Interfaces}{April 14--17, 2021}{College Station, TX, USA}
\acmBooktitle{26th International Conference on Intelligent User Interfaces (IUI '21), April 14--17, 2021, College Station, TX, USA}
\acmPrice{15.00}
\acmDOI{10.1145/3397481.3450676}
\acmISBN{978-1-4503-8017-1/21/04}

\begin{document}

\title{XAlgo: a Design Probe of Explaining Algorithms' Internal States via Question-Answering}

\author{Juan Rebanal}
\email{jrebanal@ucla.edu}
\affiliation{%
  \institution{UCLA HCI Research}
  \country{USA}
}

\author{Jordan Combitsis}
\email{jcombitsis17@gmail.com}
\affiliation{%
  \institution{UCLA HCI Research}
  \country{USA}
}

\author{Yuqi Tang}
\email{yukitang0703@gmail.com}
\affiliation{%
  \institution{UCLA HCI Research}
  \country{USA}
}

\author{Xiang `Anthony' Chen}
\email{xac@ucla.edu}
\affiliation{%
  \institution{UCLA HCI Research}
  \country{USA}
}

\renewcommand{\shortauthors}{Rebanal, et al.}

\begin{abstract}
  % Algorithms often appear as 'black boxes' to non-computing users. While prior work focuses on explainable representations and expert-oriented exploration, we propose XAlgo---a generalizable interactive approach using question answering to explain non-statistical algorithms to users without a computational background. We develop a formal model that first classifies the type of question based on a taxonomy, and provides a template whereby an answer is generated based on a set of rules that extract information from representations of an algorithm's internal states, \eg the pseudocode. A user study in an algorithm learning task with 18 participants (9 for a Wizard-of-Oz XAlgo and 9 as a control group) reports what kinds of questions people asked, how well XAlgo responds, and what remain as challenges to bridge non-computing users' gulf of understanding algorithms.

% % promise -------------------------------------------------

% % problem -------------------------------------------------

% % prior art -----------------------------------------------

% % proposed solution ---------------------------------------

% % proof ---------------------------------------------------

Algorithms often appear as 'black boxes' to non-expert users. While prior work focuses on explainable representations and expert-oriented exploration, we propose and study an interactive approach using question answering to explain deterministic algorithms to non-expert users who need to understand the algorithms' internal states (\eg students learning algorithms, operators monitoring robots, admins troubleshooting network routing). We construct XAlgo---a formal model that first classifies the type of question based on a taxonomy and generates an answer based on a set of rules that extract information from representations of an algorithm's internal states, \eg the pseudocode. A design probe based on an algorithm learning scenario with 18 participants (9 for a Wizard-of-Oz XAlgo and 9 as a control group) reports findings and design implications based on what kinds of questions people ask, how well XAlgo responds, and what remain as challenges to bridge users' gulf of algorithm understanding.

% \xac{mention the relevance to user interface}
\end{abstract}





\maketitle

\section{Introduction}

% promise -------------------------------------------------
% problem -------------------------------------------------
The world is increasingly run by automation algorithms \cite{Steiner2012}, most functioning as `black boxes' to people with limited algorithm expertise.
As a result, non-expert users are often left with no ways to comprehend how or why an algorithm arrives at certain (unexpected) results, causing a lack of transparency and explainability in algorithm-driven scenarios. 
% When algorithms become trusted to make decisions in ethically loaded situations, being able to trust them by way of them explaining themselves becomes important. 

% prior art -----------------------------------------------
To solve this problem, prior work proposed explainable representations via extracting rules to approximate an algorithm's behavior \cite{Frosst2018,Ribeiro2016-vc,Ribeiro2018}, visualizing components of an algorithm \cite{Olah2017,Olah2018}, or tracing parts of the input to attribute the algorithm's output \cite{Shrikumar2017-lq,Zintgraf2017-ey}. 
Others went beyond static representations to provide interactive tools that support visual analysis of algorithms for computing experts (\eg developers, data scientists) \cite{talbot2009ensemblematrix, may2011guiding, krause2016interacting, strobelt2016visual, kahng2016visual, krause2018using}, or to enable non-expert domain users (\eg physicians \cite{xie2020chexplain, Wang2019}) to explore and understand a system's computational model.

Most of the interactive algorithm explanations have been taking what Wick and Thompson \cite{Wick1992} called a \textit{reconstructive} approach: instead of `breaking into' a black-box model, one can reconstruct a model's behavior by visualizing, inspecting and tweaking its input and output data. The rationale is that by reconstructing a more explainable `proxy' that approximates the original model, one can allow non-expert users to obtain a level of understanding beyond a black box.

% \fg{fig_1}{fig_1}{0.8}{XAlgo is a generalizable interactive approach using question answering to explain deterministic algorithms to non-expert users. Our main contribution is a formal model that first classifies a user's question by a taxonomy, which guides our proposed model to locate relevant algorithm states and extract relevant information to compose an answer.
% \vspace{-2em}
% }

However, a reconstructive approach presents limited insight for users who do need to understand the internal mechanisms of algorithms, \eg students learning sorting functions, a network administrator troubling-shooting the system's routing policy, 
% an artist using a library to program an interactive installation, 
and a remote operator analyzing what causes a field robot's abnormal behavior. On the spectrum of expertise, such a group of people are situated between algorithm developers and end-users: although they do not have the first-hand knowledge as the algorithm developers, they do need to achieve a level of algorithmic understanding that goes beyond end-users. 

Our goal is to explore interactive mechanisms that support these users not only by showing them how algorithms work, but, more importantly, to answer algorithm-related questions asked by users. 
Asking questions is a natural way for seeking explanations in human-human communication and answering users' questions naturally presents itself as a user-centered approach towards eXplainable AI (XAI).
Specifically, as a starting point, this work focuses on explaining {\it internal states of deterministic algorithms} that do not depend on statistical training data.

To discover the potentials and problems of such a question-answering (QA) approach for algorithm explanations, we conducted a design probe:

\begin{figure}%{r}{0.37\textwidth}
      \includegraphics[width=.9\linewidth]{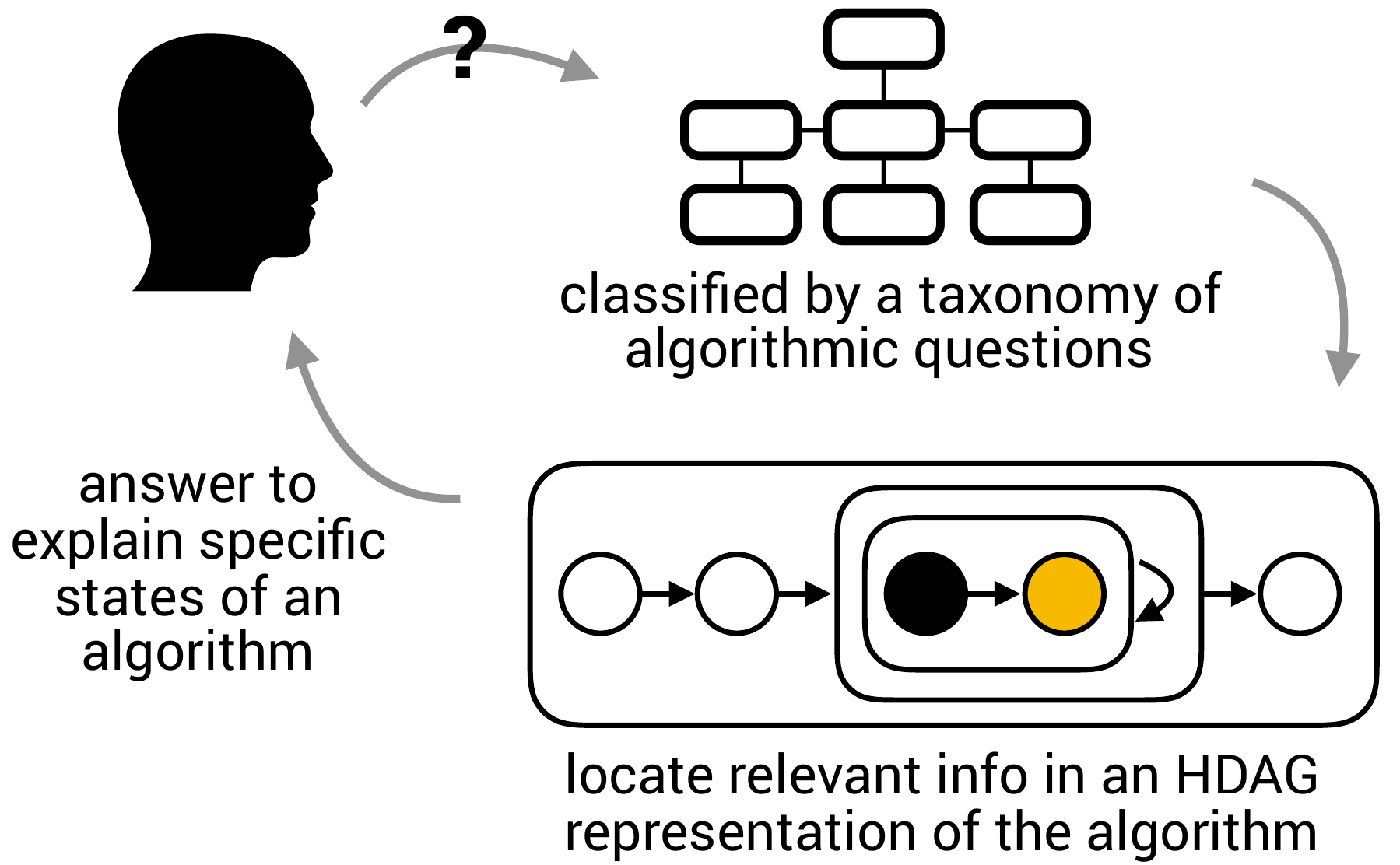}
    \caption{Our design probe employs XAlgo---a model that classifies a user's question by a taxonomy, locates relevant algorithm states, and extracts relevant information to compose an answer.}
    \label{fg:fig_1}
  \end{figure}

% 
% proposed solution ---------------------------------------
% We propose XAlgo (eXplainable Algorithm)---
First, we propose a formal model, called XAlgo, that generates answers to a user's question related to specific states of an algorithm's execution. 
% Specifically, we focus on {\it non-statistical algorithms}, \ie algorithms with self-contained logic without requiring training data.
As shown in \fgref{fig_1}, XAlgo consists of two steps: \one identifying question type---we draw from literature \cite{Miller2018, hankinson2001cause,Kass1987, Girju2003} and more specifically NLP research on question answering \cite{green1961baseball,yih-etal-2014-semantic,Hermjakob2000,Li2002,Wood} to build a taxonomy of questions one might ask (what, why, how and yes/no) in relation to understanding an algorithm; \two composing the answer---based on the identified question type, XAlgo employs a set of rules for extracting information from representations of an algorithm (\eg the pseudocode), which serve as building blocks of composing the answer. 

\begin{figure}%{r}{0.37\textwidth}
      \includegraphics[width=.9\linewidth]{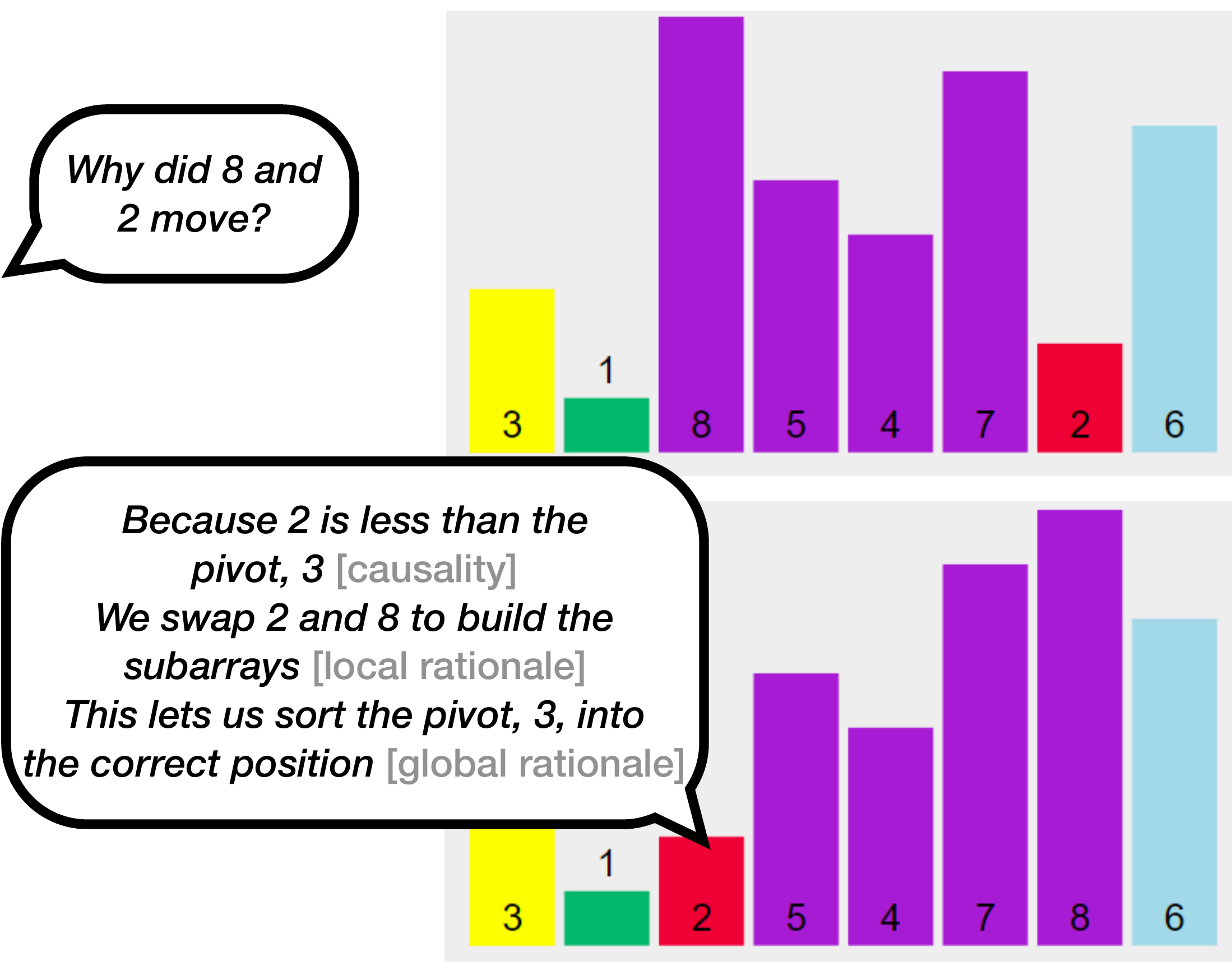}
    \caption{Our design probe is situated in an algorithm scenario where XAlgo answers a user's question about an animation of QuickSort.}
    \label{fg:scenario}
  \end{figure}

Next, 
% To ensure that we get ``the right design'' \cite{tohidi2006getting} as opposed to ``a design right'', 
we employ a Wizard-of-Oz method to instrument XAlgo in 
% an algorithm learning scenario: rather than testing users' acceptance of a system design (\eg \cite{Gould1983,Ferreira2019,Lee2019}), we study the performance of XAlgo's model (similar to \cite{Maulsby1993,Chang2019,Jain2019})---\ie whether the model can answer users' questions to help them understand an algorithm. 
% 
% \subsection{Example Scenario: XAlgo for Learning Algorithms}
% We instantiate XAlgo in 
one of the aforementioned scenarios---a tutoring system for students to learn sorting functions---as a Wizard-of-Oz platform on top of VisuAlgo\footnote{A website that curates animations of data structures and algorithms. \url{https://visualgo.net/en}}.
\fgref{scenario} shows a snapshot of a QuickSort algorithm \cite{hoare1962quicksort} where one of our study participants asked ``Why did 8 and 2 move?''. XAlgo classifies this question as both Causality and Rationale related, and composes the following answer (annotations in brackets are added for illustration purposes):
``{\it Because 2 is less than the pivot, 3.} [causality]~~~
{\it We swap 2 and 8 to build the subarrays.} [local rationale]~~~
{\it This lets us sort the pivot, 3, into the correct position} [global rationale]''
The first sentence explains causality---that 2 being smaller than the pivot 3 causes the algorithm to swap 2 and 8. The second sentence explains the `local' rationale---the immediate goal where sorting 2 and 8 is to separate elements into two subarrays: one smaller and the other greater than the pivot 3. The third sentence explains a `global' rationale by going one level higher---by having the subarrays sorted we can then put the pivot at the correct position (between the subarrays).

% proof ---------------------------------------------------
Finally, based on this algorithm learning scenario, we conducted a design probe with 18 participants\footnote{All with < 1 year programming experience and < 3 college-level programming courses.} (9 for a Wizard-of-Oz XAlgo and 9 for a control group). Specifically, the wizard followed XAlgo's model to manually generate answers on-the-fly to participants' questions.
Results show that
\begin{enumerate}
     \item Participants asked a wide variety of questions covering most parts of the taxonomy, although nearly half the questions were concept-related that did not directly contribute to their understanding of the algorithm; 
     \item Participants found that XAlgo's answers provided useful and accurate information that grounded their understanding of the algorithm, although formulating questions, algorithmic vocabulary, and processing information in the answers remained the three major challenges;
     \item Neither group's participants performed well in quizzes that tested their learning of the algorithm, suggesting that for learning, more active guidance is needed beyond (passively) waiting to answer users' questions. 
\end{enumerate}
% \one participants asked a wide variety of questions covering most parts of the taxonomy (\fgref{fig_question_classification}); 
% \two participants found that XAlgo's answers provided useful and accurate information that grounded their understanding of the algorithm, although formulating questions, algorithmic vocabulary and information processing remained the three main challenges;
% \three neither group's participants performed well in quizzes that tested their learning of the algorithm, suggesting that for learning, more active guidance is needed beyond (passively) waiting to answer users' questions.

% We close by discussing the technical feasibility of implementing XAlgo, as well as design recommendations for future development of Explainable AI systems.

% \fg{scenario}{scenario}{0.9}{One of the many scenarios XAlgo can be useful for is learning algorithms, \eg answering a user's question about an animation of a sorting algorithm}

% \subsection{Contributions}
% Our main contributions are as follows:
% \begin{itemize}
%     \setlength\itemsep{0em}
%     \item Conceptual contribution: an interactive question answering approach for explaining the internal states of deterministic algorithms; 
%     \item System design contribution: a formal model that first identifies the type of algorithm-related question and then composes answers based on existing representations of the algorithm, \eg pseudocode;
%     \item A Wizard-of-Oz study that validates the formal model and summarizes further design and implementation recommendations.
% \end{itemize}

Our main contribution is findings from a design probe that summarize important lessons learned from instrumenting a question-answering mechanisms to explain an algorithm's internal states, as well as design implications for future QA-based XAI systems.

\section{Related Work}
To inform the design of a question-answering approach for explaining algorithms, we first review two key related bodies of work: explainable AI and question-answering research. We further summarize prior work on understanding software---a similar objective but a different audience (developers); and past research on intelligent tutoring systems---a similar approach (learning) to achieve the understanding of algorithm.

\subsection{Explainable AI (XAI)}
Foremost, our work is motivated by a lack of explainability in `black boxes' AI-enabled systems. A large body of recent work has focused on making data-driven AI models explainable, as the representations of such models (\eg a neural network) often do not permit a user to understand how a model works, why it works, or why it does not work. Ras \etal~ summarize three families of XAI methods \cite{ras2018explanation}: \one {\it rule-based}, where rules are extracted that match how a `black-box' model process certain input to produce certain output (\eg \cite{Frosst2018, ribeiro2016should, Ribeiro2018}); \two {\it attribution}, where certain output is attributed to specific parts of the input or specific components of the model (\eg \cite{Simonyan2013, Olah2017, Dabkowski2017, Olah2018,Park2018}); and \three {\it intrinsic methods}, elements of a model that are intrinsically explainable (\eg \cite{Zheng2019}) without adding any rules or attribution . Doran \etal~ review AI-enabled systems with a spectrum of explainability, ranging from opaque, to interpretable, to comprehensible, and to (truly) explainable \cite{Doran2018}. Hoffman \etal discuss metrics to evaluate XAI, including the goodness of explanations, whether users are satisfied by explanations, how well users understand the AI systems, how curiosity motivates the search for explanations, whether the user's trust and reliance on the AI are appropriate, and how the human-XAI work system performs \cite{Hoffman2018}. Miller conducted a comprehensive review on XAI in connection with relevant disciplines, \eg communication, cognitive science and sociology \cite{Miller2019}.

The goal of our paper differs the prior work above: rather than explaining data-driven AI (\eg neural networks), we focus on process-driven algorithms: whereas the former approach is based on a statistical model parameterized by training data, we are interested in the latter, where an algorithm itself self-sufficiently describes the logic of solving a problem. Although algorithms already have a representation of the problem-solving process, such representations (\eg pseudocode, block diagram, finite state machine) remain obscure to people without a computational background. Wilhelme \etal propose a shape analysis technique that visualizes key states and structural properties of a data structure (\eg a heap) during the execution of an algorithm \cite{Wilhelm2002}. Shakshuki \etal develop SHALEX---a system for explaining algorithms that addresses a lack of multi-level abstraction and user interaction in prior algorithm visualization \cite{Kerren2006}; a follow-up paper further provides users with the ability to write their own explanations for events as a way of learning \cite{Shakshuki2007}. One important observation related to all this work, as pointed out by Miller, is that explanation is by nature social---``{\it a transfer of knowledge, presented as part of a conversation or interaction}'' \cite{Miller2019}. Below we switch our attention to one of the most relevant form of conversation that can potentially facilitate explanation--question answering.

\subsection{Question Answering (QA)}
Question answering (QA) intersects natural language processing and information retrieval. The process of achieving QA can be decomposed into two steps: interpreting a question and generating an answer. Below we sample a selected amount of prior work to illustrate the background of QA research.

\insec{Classifying a question} involves associating a question with a category or a topic where there is sufficient knowledge base for answer generation. Mohasseb \etal take a grammar-based approach to classify a question into one of the six categories: confirmation, factoid, choice, hypothetical, causal, and list \cite{Mohasseb2018}. However, such a static categorical label is topic-invariant and does not provide insight of a question's actual content, \ie which topic the question-asker is interested in. To classify the topic of a question, prior work has taken a language-centric approach, \ie following rules and models of human language to parse the question. Hermjakob \etal develop `QTargets'---a type of question-answering classes, and employed a hierarchical tree-like model for parsing a question based on specific components of the speech \cite{Hermjakob2000}. Li \etal take a language modeling approach and classify questions by determining whether certain questions belong to the same dialogue \cite{Li2002}. Girju \etal focus on the extraction of causal relations in a question by keywords and develop three types of `causatives': simple, resultative and instrumental \cite{Girju2003}. In the meantime, complementary to using a language model, researchers also take a data-driven approach for question classification. Others perform coarse-to-fine classification of the topic associated to a question \cite{Li2002, Huang2008}. Beyond question classification, there is another school of NLP work focused on semantic parsing to a specific structure or format \cite{berant-etal-2013-semantic,artzi-zettlemoyer-2013-weakly,xu2018sqlnet}.

\insec{Generating an answer} is often considered as an information extraction (IE) problem, which has already accumulated a large body of prior research. For example, Yih \etal consider `single-relation question', \eg ``When were DVD players invented?''; the answer is an entity that can be connected to another entity in the question (`DVD players') by a relation (be-invented-in) \cite{Yih2014}. Nie \etal augment community QA, using textual answers provided by an online community to gather relevant multimedia information on the web to enrich the answer \cite{nie2017data}. 
The advent of data-driven learning gives rise to new QA development based on large corpora of examples \cite{rajpurkar2016squad,rajpurkar2018know}.
Our research on XAlgo analogously augments existing animated illustration of algorithms by extracting information at each step to answer a user's context-dependent questions. Indeed, the use of visuals is often inseparable from QA: Visual QA uses a neural model to recognize relevant entities in an image and uses that information to compose an answer \cite{Antol2015}. Specifically, we are interested in extracting answers to algorithm-related questions, which intersects two major application domains: support for understanding software and tutoring technology based on chatbots, which we review below.

% Like my comments above, I'm not quite sure what is the goal of literature survey for this paragraph.

% I'm not sure if you're trying to say there are different types of QA problems in which the question and the answer can be in different forms. 

% Or you're trying to say there are different techniques have been studied for answering questions. 

\subsection{Understanding Software Codes \& Programs}
Algorithm explanation is often manifested as source code exploration and understanding in software engineering. For example, Alber \etal employ an attribution-like method to explain an image classification program by attributing a result to specific parts of a software project \cite{Alber2019}. Oney \etal further incorporate communication as a means to facilitate explanation of source code, providing integrated support for instructors and learners to chat with one another with directed references to the source code \cite{Oney2018}. To generate explanation-specific answers about a program's behavior, Ko \etal retrieve a causal chain---a flow of actions that lead the specific outcome questioned by the user \cite{Andrew2004}. Wood \etal collected a corpus of QA conversations during bug repair and distill a set of ``speech acts'' that signal specific tasks in question \cite{Wood}. Different the scope of our research, all this work focuses on professional software developers; for users without a computational background, being explained about an algorithm is almost indistinguishable from learning that algorithm. Below we briefly review intelligent tutoring systems, focusing on systems that employ chatbots.

\subsection{Intelligent Tutoring Systems using Chatbots}
The use of chatbot-like natural language based interaction is widely used in building intelligent tutoring systems. AutoTutor is a dialogue-based tutor that provides conversation-facilitating features, \eg giving different good/bad/neutral feedback, prompts, ``pumps'' (inviting the learner to speak up), and hints to promote an interactive learning experience \cite{Graesser2004}. Van Lehn \etal find that question-answering serves as an effective means to fill the granular `gaps' between the student and the tutor \cite{VanLehn2011}. Fei \etal find that chatbots as a communicative approach to learning allows learners to study in low-anxiety situations and allow learners to engage material when they feel ready: students are more willing to repeat themselves/ask chatbots to repeat themselves \cite{Fei2013}. Ruan \etal find that students strongly prefer using a chatbot to learn and memorize factual knowledge compared to traditional flash cards \cite{Ruan2019}. Cai \etal develop a rule-based chatbot for learning math and attain a strong preference when employed in an online video tutorial system \cite{Cai2019}.

\section{Formal Model}

XAlgo responds to an explanation-seeking question by first classifying the question into one of the five major categories (\fgref{fig_question_classification}), and then generating the answer using a set of rules to extract  information from the relevant internal states of the algorithm (\fgref{fig_hdag}). Below we detail the model's process of classifying questions and generating answers.

\subsection{Question Classification}
% related work
Question classification is a well-established problem in NLP (\eg \cite{chisholm1984interrogativity, green1961baseball, Hermjakob2000, li2002learning, Li2002, Huang2008, yih-etal-2014-semantic, Mohasseb2018}). As shown in \fgref{fig_question_classification} and detailed in this subsection, the main difference of our model is the last `layer' that follows the initial {\it wh}-word based classification, where we taxonomize five categories of explanation a question might be seeking.

As a user asks a question about the algorithm, the model first extracts the following information required for the subsequent classification of the question:

\begin{itemize}
    \item \insec{interrogativeWord:} also known as the \textit{wh}-word, \eg why, what, how. Such words usually occur at the beginning of a question and are used to identify the overall question type;
    \item \insec{timeShift:} whether the user is asking about past, now or future states of the algorithm based on words \eg next, last, after, as well as grammatical tense. 
    % If no such words are detected, the user is assumed to be talking about the current state (that the question is asked in);
    \item \insec{objects:} algorithm entities, \eg elements in an array;
    \item \insec{values:} parameters of \insec{objects}, \eg the number carried by each array element;
    \item \insec{action:} what the algorithm does with the \insec{objects}, \eg swapping two elements in an array;
\end{itemize}

Next, the model filters questions that are concept related. In the informal pilot studies, we noticed a subset of questions that were related to the concepts of the algorithm, rather than to the specific states of the ongoing algorithmic process. For example, ``What is this algorithm trying to do?'', ``What is a pivot?'', and ``Why is that number selected as the pivot?'' Currently we answer such concept-related questions using a manually one-time generated look-up table, although more advanced, automatic methods can be used (\eg based on mining code comments \cite{wong2015clocom}), which we leave as future work.

If the question is not concept-related, the model then classifies the question based on its interrogative word as one of the five question types as shown in \fgref{fig_bigtable}.

% \cleardoublepage

% \fgw{fig_bigtable}{fig_bigtable}{1.0}{\xx}

\subsection{Answer Generation}
Based on the identified type of question, the next step is to locate relevant internal states of an algorithm and to extract relevant information from that state to generate an answer. We first introduce a general representation XAlgo uses to model and traverse an algorithm's states.

\fgw{fig_question_classification}{fig_question_classification}{0.85}{
    XAlgo classifies an algorithm-relate question by first extracting key information and determining whether the question is asking about concept (that can be answered by looking up a table of pre-generated concepts); if not, the question is classified first by its interrogative word, and finally into one of the five question types, which are illustrated with detailed examples in \fgref{fig_bigtable}.
    \vspace{-1em}
}

% an answer template is produced. Next, the model fills in the template by locating and extracting relevant information from the representations of an algorithm.

{\bf Hierarchical Directed Acyclic Graph}. 
Pseudocode is a commonly used representation for algorithms, as it concisely and universally describes the logic of the algorithm independent of the algorithm's implementation or the top-level application. 
To create a more convenient way to look up information in the pseudocode, we employ a data structure akin to the Hierarchical Directed Acyclic Graph (HDAG) \cite{suzuki2003hierarchical}. Essentially, an HDAG transforms pseudocode into a state diagram with hierarchy. The rules of creating an HDAG are as follows:

\begin{itemize}
    \item  
    To create hierarchies: an algorithm starts as a (root) DAG, and each loop, conditional branch or recursive call creates a child DAG. In this way, a DAG can be thought of as a high-level operation, which is manifested as branches or iterations of executions under its hierarchy; conversely, a low-level operation (\eg a node) is being governed by or is contributing to the parent DAG's operation.
    % Each iteration of a loop can be thought of as performing a higher-level operation, and each such higher-level operation consists of several lower-level operations, manifested as the individual lines of code;

    \begin{figure*}
        \centering
        \includegraphics[width=\textwidth]{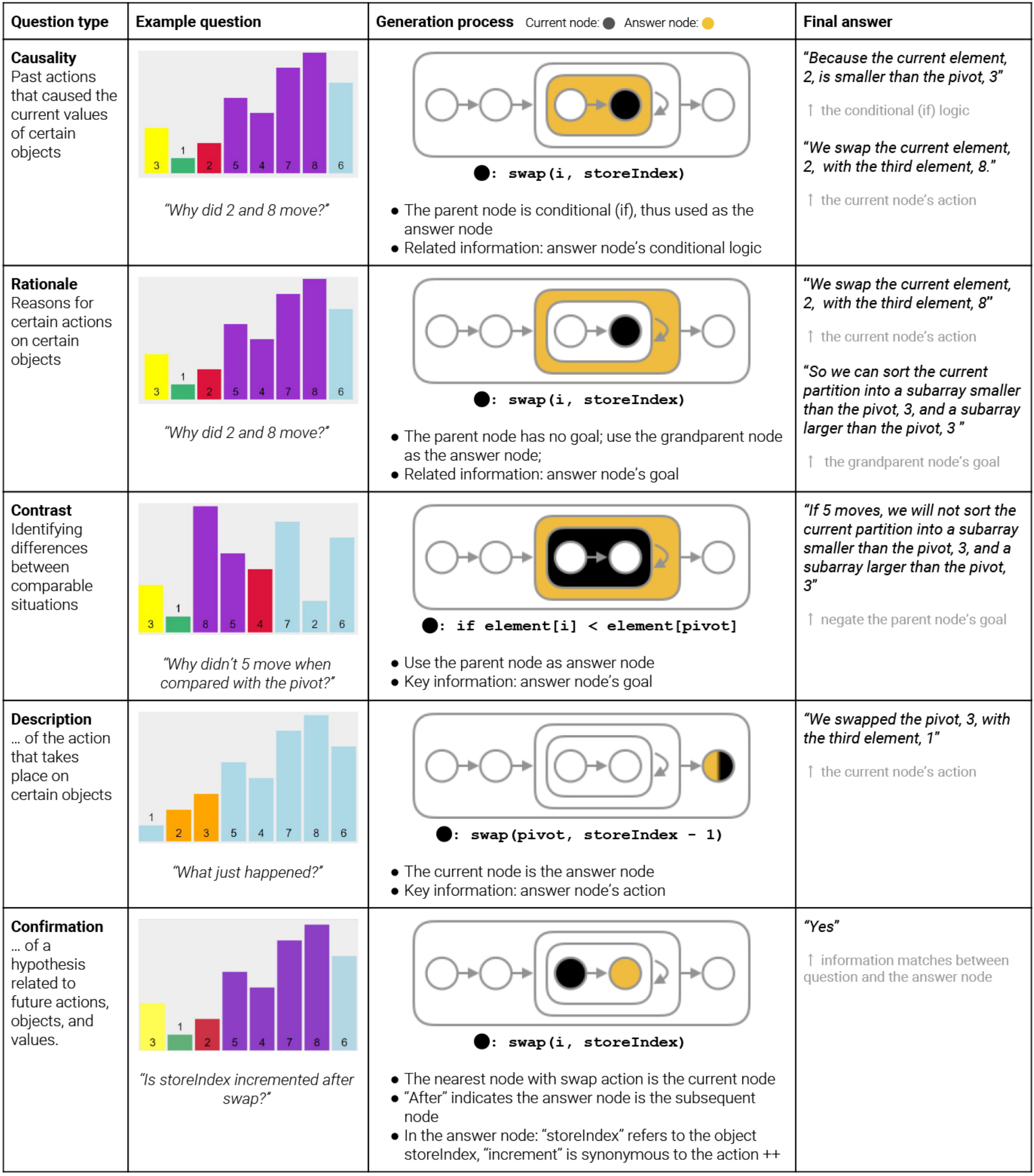}
        \caption{Examples of answer generation for each question type classified from \fgref{fig_question_classification} based on the VisuAlgo animation of a QuickSort algorithm.}
        \label{fg:fig_bigtable}
    \end{figure*}

    \item 
    To create nodes: if a line of pseudocode causes a value change of objects in the main data structures, we create a new node (however, other changes such as iterator i++ does not warrant new nodes). The idea is that each node corresponds to a perceivable intermediate state of an algorithm, \eg array elements changing positions while being sorted. In this way, a question asking about the algorithm at the current state can be mapped to a specific node for answer generation.
    
\end{itemize}

% \pagebreak

% \fg{fig_hdag}{fig_hdag}{1.0}{An exemplar Hierarchical Directed Acyclic Graph (HDAG) constructed from the pseudocode of the QuickSort algorithm: each loop or conditional block creates a new hierarchy whereas each line of code can be represented as a node that contains objects (\eg an array), values (\eg numbers stored in an array), action (\eg swapping two array elements), and goal (a description of what this node does).
% }

\fgref{fig_hdag} shows an exemplar HDAG representation of the QuickSort algorithm on a simple array for illustration purposes. Importantly, besides the basic algorithmic constructs (\insec{object}, \insec{value}, \insec{action}), each node should also contain a description of its \insec{goal}, which can be annotated as a one-time preprocessing step, similar to how programmers comment their code. For example, the for-loop in QuickSort (\fgref{fig_hdag}) has a goal of ``sorting the pivot in the right place'', and the if-statement has a goal of  ``compare the current element and the pivot''.

% \missingfigure{QuickSort algorithm on a simple array for illustration purposes}

{\bf Locating the answer node}
% A question is asked during the execution of an algorithm, which is represented as an HDAG. 
Based on the HDAG representation, we first identify the current node associated with the current state which the question is asked about. Specifically, we use \insec{timeShift}---if there is any---to `shift' to the antecedent or subsequent nodes. For example, as shown in \fgref{fig_bigtable}, the question ``Is storeIndex incremented after swap?'' causes a shift to the subsequent node.
We then continue to locate the `answer node'---the actual node where the answer can be generated. If the question is Description or Confirmation, the answer node is the current node. 
\fgref{fig_bigtable} shows several specific examples.

\begin{figure}%{r}{0.5\textwidth}
      \includegraphics[width=0.48\textwidth]{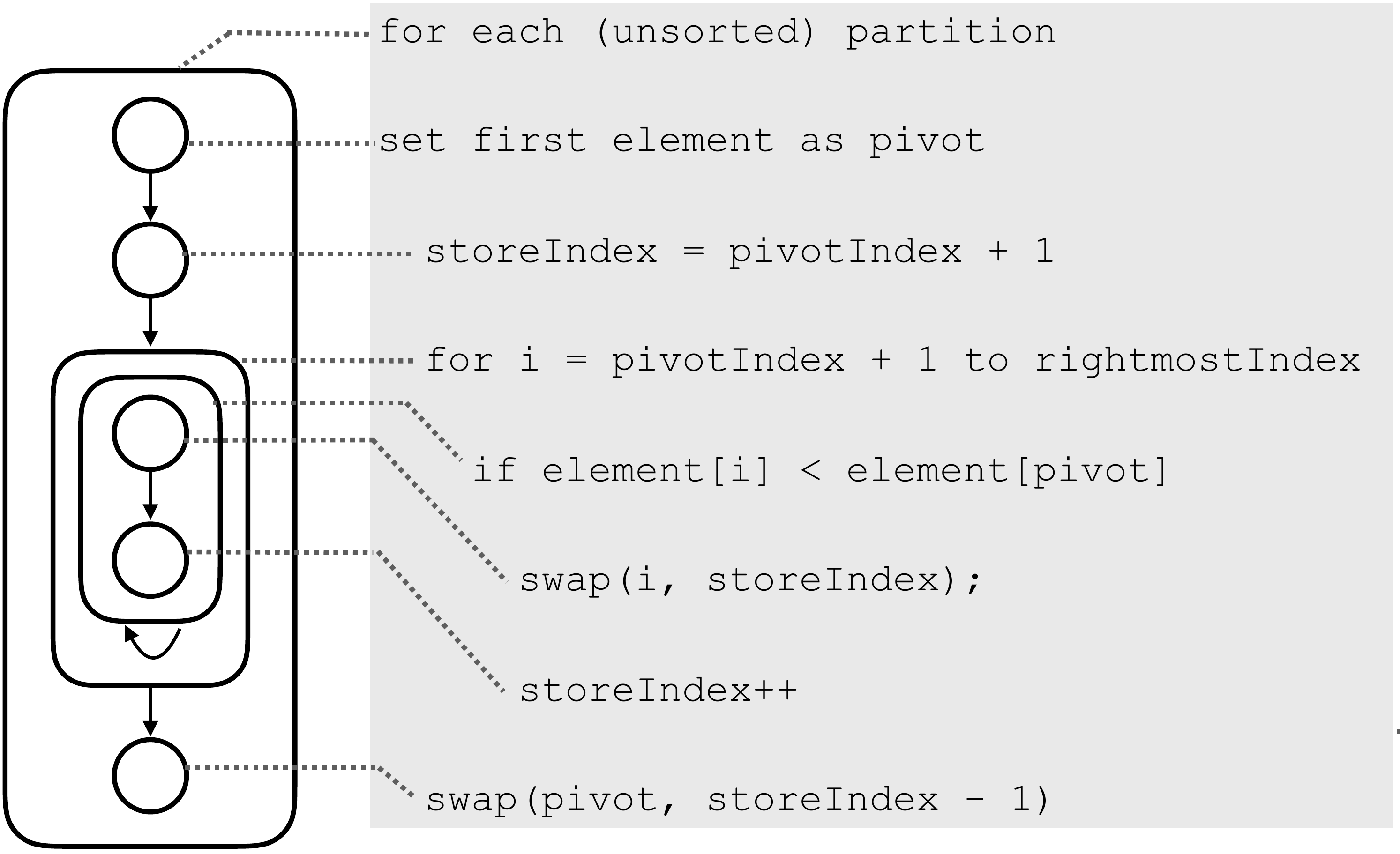}
    \caption{An exemplar Hierarchical Directed Acyclic Graph (HDAG) constructed from the pseudocode of the QuickSort algorithm: each loop or conditional block creates a new hierarchy whereas each line of code can be represented as a node that contains objects (\eg an array), values (\eg numbers stored in an array), action (\eg swapping two array elements), and goal (a description of what this node does).}
    \label{fg:fig_hdag}
  \end{figure}

If it is a Causality question, the model considers two cases: if the node is nested in a conditional statement, we use the node that contains that statement as the answer node; otherwise we use the current node's antecedent as the answer node. For example, as shown in \fgref{fig_bigtable}, ``Why did 2 and 8 move'' finds the answer node as its parent node, which is an if-statement (see corresponding source code in \fgref{fig_hdag}).
If it is a Rationale or Contrast question, we use the current node's parent as answer node. For example, as shown in \fgref{fig_bigtable}, the same question ``Why did 2 and 8 move'' can also be answered with a rationale, which is the goal of its grandparent node.
% \xac{}{maybe: example}
It is possible that a user might ask a `context-independent' question, \ie the answer to which is unrelated to the current node of algorithm execution. In such cases, we find the nearest node that matches a user's question. 
% \xac{}{give a concrete example here}

% the first step of processing the question is to shift the reference state from which all answers will be generated from the state the question is being asked to the state the question is asking about.

{\bf Extracting information to compose an answer}. Once the answer node is located, we employ the following rules to generate answers. If a question is Causality, Rationale or Description, we simply describe the answer node's \insec{goal} (with some natural language components to address the specific question type, \eg 'because', 'so', 'if'). If a question is Contrast, we negate the answer node's \insec{goal}. If a question is Confirmation, we compare the \insec{object}s, \insec{value}s and \insec{action} between the question and the answer node, and inform the user whether their hypothesis matches the referred state in the algorithm.

The last column in \fgref{fig_bigtable} shows examples of generated answers. Note that within the scope of this paper, all the natural language components were manually added, which can be further automated as a post-processing step. We leave this as future work.
\section{Design Probe}
To validate XAlgo's model without the cost of system implementation, we conducted a Wizard-of-Oz study of XAlgo in an algorithm learning application scenario. Specifically, we sought to answer the following research questions:

\begin{itemize}
    \setlength\itemsep{0em}
    \item RQ1. What kinds of questions do users ask XAlgo?
    \item RQ2. How well can XAlgo answer users' algorithm-related questions?
    \item RQ3. How well can users achieve the application-specific task, \ie learning an algorithm, using XAlgo?
\end{itemize}

To investigate these questions, we situate XAlgo in an algorithm learning/understanding scenario based on the aforementioned VisuAlgo platform (\fgref{study}). We chose QuickSort as the target of explanation, as it is a commonly used algorithm. Compared to `entry-level' sorting algorithms (\eg Bubble Sort), QuickSort has a certain amount of complexity that could elicit explanation-seeking questions from users trying to learn the algorithm.

\subsection{Design \& Participants}
We employed a between-subject design. The independent variable was {\it System} (XAlgo's QA+animation vs. VisuAlgo's animation-only).
We recruited 18 participants--9 for each condition--via convenience sampling, flyers in a local university and a Craigslist online posting. We employed a screening questionnaire as people signed up for the study. Specifically, we only accepted participants with none or less than one year of programming experience, who had taken fewer than three college-level computer science courses. Specific to the algorithm chosen for the study, we also made sure participants had no or limited understanding of sorting (by asking them to describe QuickSort in the screening questionnaire).
Amongst the 18 participants, there were 8 male, 10 female, aged 19 to 25. None of the participants had studied or was studying any fields related to Computer Science or Electrical Engineering.
Participants were randomly assigned to one of the two conditions.
Each participant received a \$25 gift card as compensation.

\subsection{Apparatus \& Setup}
\begin{figure}
\includegraphics[width=1\columnwidth]{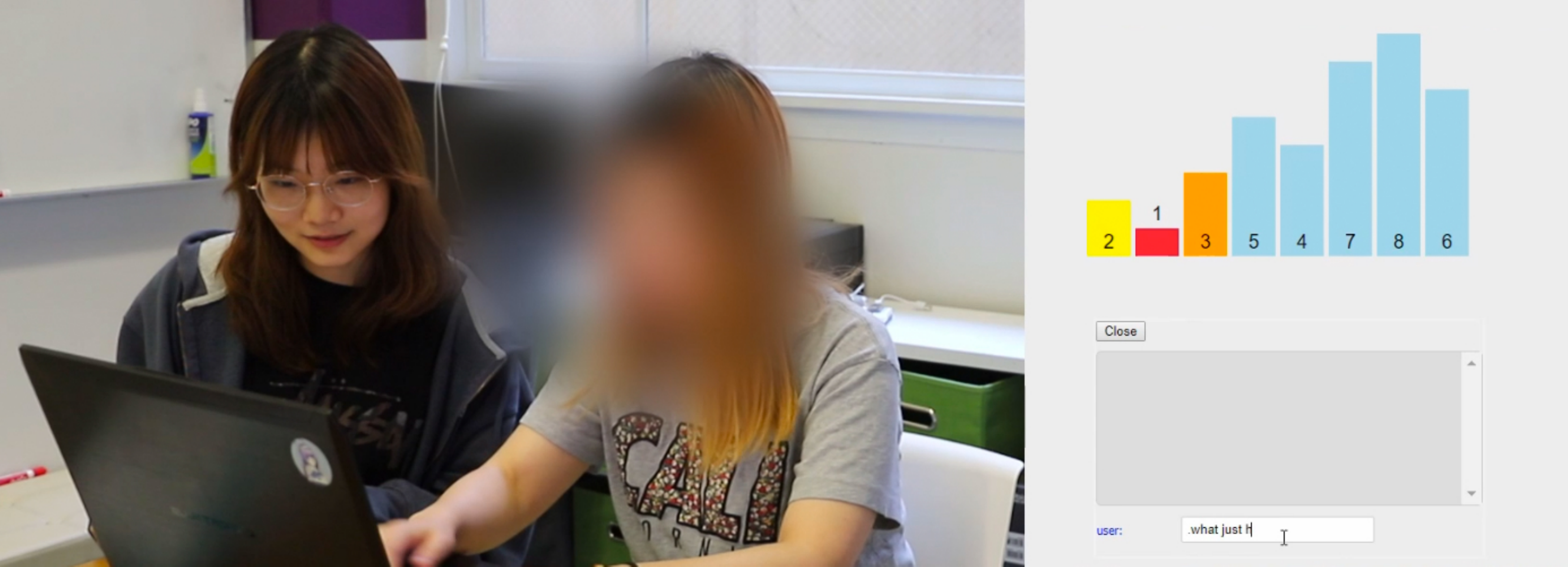}
\caption{We conducted a Wizard-of-Oz study of XAlgo deployed on the VisuAlgo platform. The second experimenter (the wizard) was in the same room but not shown in this image.}
\label{study}\end{figure}

To implement XAlgo's model, we used VisuAlgo's pseudocode to pre-populate an HDAG with one-time annotations for answer generation. To implement XAlgo as a test platform, we modified the VisuAlgo codebase by adding a custom dialog box (\fgref{study}). This allowed for the use of VisuAlgo's visualizations and pseudocode to naturally prompt the user to ask questions about the algorithm. We also added a communication protocol so that two XAlgo programs can talk to each other over a local area network, thus enabling the wizard's performance of XAlgo to a user. For the VisuAlgo group, we ran the unmodified VisuAlgo codebase locally.
%  on the experiment laptop.

The experiment took place in the conference room of our research lab. One experimenter guided the participant; for the XAlgo group, another experimenter played the wizard, who was in the same room using a different laptop to chat with the participants. XAlgo participants were told that the wizard was taking notes and to provide technical support when needed. 

Each  participant's screen was recorded during the entire study using Open Broadcaster Software. Meanwhile, audio was recorded using the laptop's microphone.

\subsection{Tasks \& Procedure}
\subsubsection{Introduction \& training (5 min)}. We started by introducing the common background and motivation, which was to provide an interactive environment for people to learn and understand algorithms. For the XAlgo group, we explicitly instructed participants to ask questions in order to understand the algorithm. Next, we provided a hands-on tutorial of the system (VisuAlgo or XAlgo)and let each participant try out the interaction.\enlargethispage{13pt}

\subsubsection{Interacting with the system \& multiple-choice quizzes (30 min)}. Once participants familiarized themselves with the interaction, they were asked to interact with the system, trying to learn and understand QuickSort by watching how an array was sorted (VisuAlgo) and also by asking questions (XAlgo). Participants were given six multiple-choice quiz questions they must finish before the session ends.
% Participants are not allowed to change their answer once they made their choices. 
% We chose sorting algorithms as they were one of the foundational topics in learning algorithms; we chose QuickSort as it is one of the most commonly used examples for teaching sorting algorithms.
% 
In the XAlgo group, upon receiving a participant's question, the wizard instantly responded with ``Processing ...'' and began generating an answer on the fly following the aforementioned model.
In the VisuAlgo group, both experimenters did not answer any of participants' algorithm-related questions.
In the six multiple choice questions, a participant was given a subsequence of VisuAlgo animation (of a different array) and asked to predict what was the next step. 
% Participants can start the quiz whenever they are ready, if they encounter any problems, they can turn back to the interaction website to learn. 
% 
We capped this phase to 30 minutes, although participants could choose to end early.

\subsubsection{Walkthrough quizzes (10 min) \& interviews (15 min)}. In the next phase, participants were asked to walk through a complete process of sorting a simple four-element array by filling in the blanks of the key steps and describing how the array was updated along the way. 
% During the quiz, participants were not allowed to refer to the interaction system. 
Participants were given a maximum of 10 minutes to finish the quiz. 

Then, we conducted a semi-structured interview where the participants evaluated the difficulty of the learning task, how much more they felt they understood the algorithm after interacting with VisuAlgo/XAlgo, and whether they felt the learning process was enjoyable.

For the XAlgo group, we further asked participants about whether they could understand the generated answers, whether the answers were natural, detailed and accurate, and whether the answers helped them learn and understand the algorithm. We also asked participants to comment on other positive aspects of the overall experience and areas for improvement.

\section{Findings \& Design Implications}

% \subsection{Summary of Findings}
% We provide a summary of our findings before discussing the details.
% \begin{enumerate}
%      \item Participants asked a wide variety of questions covering most parts of the taxonomy: nearly half the questions were concept-related that did not directly contribute to their understanding of the algorithm; 
%      \item Participants found that XAlgo's answers provided useful and accurate information that grounded their understanding of the algorithm, although formulating questions, algorithmic vocabulary and information processing remained the three major challenges;
%      \item Neither group's participants performed well in quizzes that tested their learning of the algorithm, suggesting that for learning, more active guidance is needed beyond (passively) waiting to answer users' questions.
% \end{enumerate}

\subsection{What questions non-expert users asked XAlgo}
During the study, the wizard followed the aforementioned classification model (\fgref{fig_bigtable}) to label the question type. After the study, the other experimenter reviewed each participant's question and, without seeing the wizard's label, perform another question type classification. Then the two experimenters resolved the differences via discussion. 
One change we made was merging the categories of Rationale and Causality questions, as we found in the informal pilot study that combining answers to both types of questions provided a more comprehensive explanation that covers both `how come' and `what for' \cite{Miller2019}. \fgref{fig_bigtable} shows such an example.\enlargethispage{13pt}

Overall, XAlgo participants asked a total of 92 questions. Four were excluded as they were not algorithm-related (\eg ``What should I do?''), leaving 88 questions answered by XAlgo---almost an average 10 questions per person. To our surprise, amongst these questions 37 (42.0\%) were concept-related, which was answered using a pre-generated look-up table. 
% (although more advanced, automatic methods are possible, which we leave as future work). 
The look-up table encompassed all the conceptual topics (\eg the concept of a pivot) but did not anticipate all the possible questions. For example, while the table contained an answer to ``What is a pivot?'', it did not provide a direct response to questions such as ``Does the pivot always have to be number in the middle?''; in such cases, the wizard would simply answer the question with the closest information from the table.

Amongst the other 51 (58.0\%) questions related to the algorithm's states, 21 (23.9\%) were Causality/Rationale, 17 (19.3\%) were Confirmation, 12 were Description (13.6\%) and only one was Contrast.
The high concentration of concept-related questions was likely because participants were unfamiliar with algorithmic terms. Unfortunately, learning these terms would not be enough to understand an algorithm; rather, it should be the XAlgo-type of `operational' questions that actually reveal an algorithm's process to a user. Further, it was a little surprising that only one question was Contrast---the reason might be participants were still at the novice phase and were not able (or not confident) to develop a contrastive hypothesis to `challenge' the algorithm.

{\bf Design implications:}
\one For non-expert users, QA-based XAI systems should be prepared to answer both algorithm state related questions and concept-related questions, the latter of which requires extra sources of information beyond an algorithm's source code or model parameters;
\two Non-expert users ask different types of questions disproportionally, where more data could be collected to compute the prior probabilities for each question type to guide the implementation of the question interpretation part of a QA-based XAI system.

% We summarize the recurring themes to outline key factors that contribute to XAlgo's ability to answer users' algorithm-related questions, as well as challenges for future work.

% \fg{fig_xalgo_feedback}{fig_xalgo_feedback}{0.6}{
%      Participants' responses to XAlgo's answers. Overall participants reacted positively to XAlgo's ability to answer their algorithm questions: the most agreed-upon aspect is providing sufficient details whereas naturalness has the most split ratings.
% % 
% }

% fig_learning_experience

\begin{figure*}
     \centering
     \begin{subfigure}[b]{0.425\textwidth}
         \centering
         \includegraphics[width=.95\textwidth]{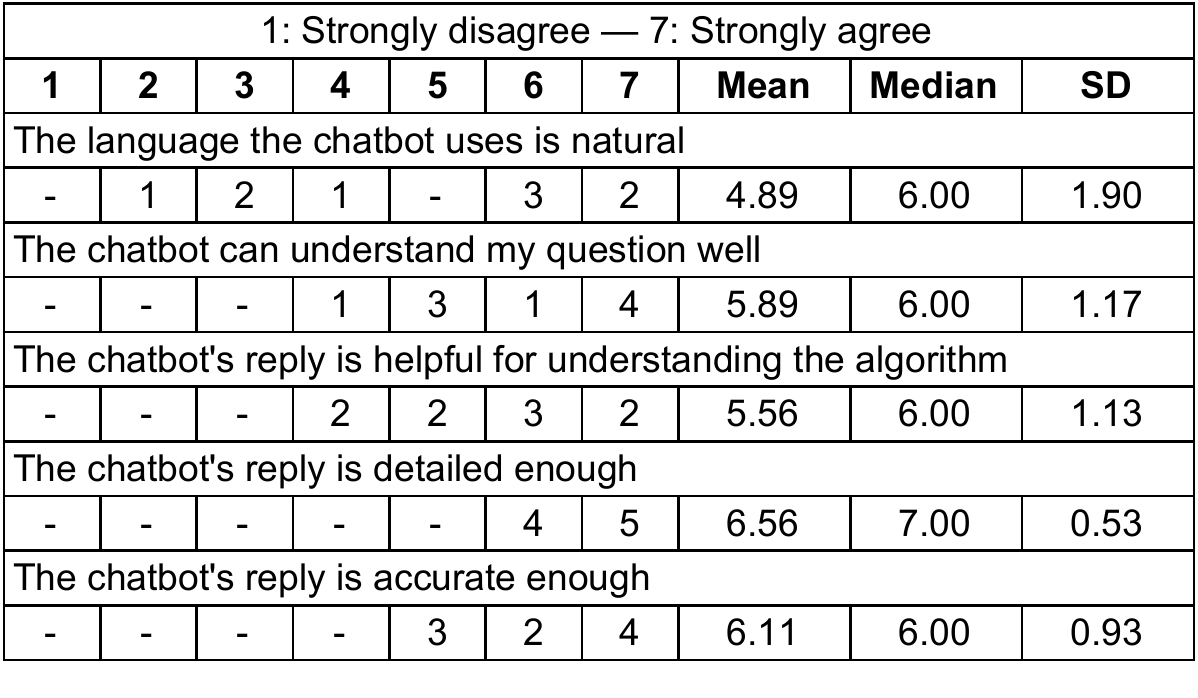}
         \caption{The most agreed-upon aspect is providing sufficient details whereas naturalness has the most split ratings.}
         \label{fg:fig_xalgo_feedback}
     \end{subfigure}
     \hfill
     \begin{subfigure}[b]{0.55\textwidth}
         \centering
         \includegraphics[width=.95\textwidth]{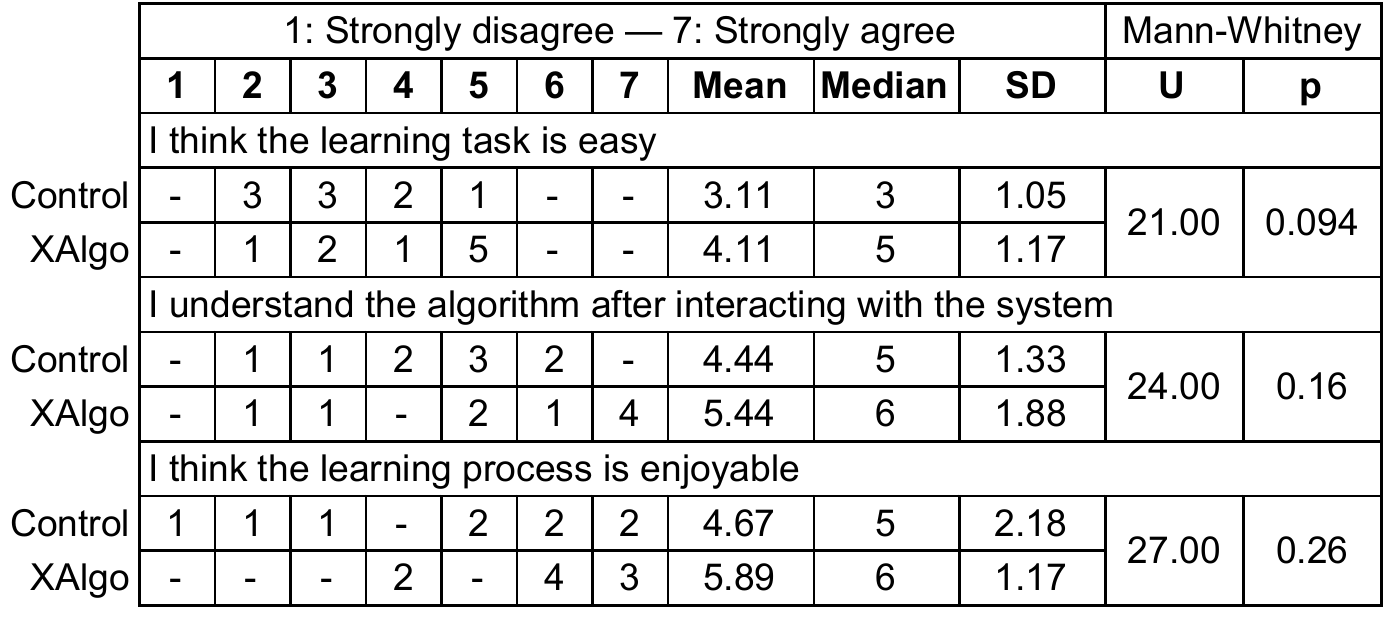}
         \caption{In all three questions, XAlgo scored at least one point higher; the differences, however, were not statistically significant.}
         \label{fg:fig_learning_experience}
     \end{subfigure}\vspace*{-6pt}
     \caption{Participants' overall responses to XAlgo's answers (a) and the learning experience (b).}
     % \label{fig:three graphs}
    %  \vspace{-2em}
\vspace*{-6pt}\end{figure*}

\subsection{How well XAlgo answers users' algorithm questions}
\fgref{fig_xalgo_feedback} summarizes participants' responses that evaluated the answers generated by XAlgo (referred to as 'chatbot' in the questions).
To obtain further insight behind these ratings, we employed an iterative open-coding method to analyze the qualitative data collected from the semi-structured interview that accompanied these questions. Two experimenters coded each participant's data one day after the study. One experimenter performed the first pass of coding and updated a codebook, which was then reviewed by the other experimenter to discuss and resolve disagreements. The two experimenters alternated the roles of first coder and reviewer. After all participants' data was coded and consolidated, a third experimenter reviewed all the codes and transcripts, and resolved disagreements through discussion with the previous two experimenters.
The main recurring themes are as follows.

\subsubsection{Groundedness}
Asking question externalizes the internal obstacle a person experiences in understanding a concept. One main benefit of XAlgo was helping participants ground their understanding of an algorithm to a specific state that gave rise to a question, \eg
% 
% \quo{... just ask at the step you're at and then like it will know like what's that you're referring to}{5}
% 
\inquo{... sometimes I just like assume I know something while I don't; so this like explains very well.}{5}

% In the design of XAlgo, a mentioning of an algorithmic term is followed by its specific value at the state where the question is asked, \eg ``we swap the pivot, {\bf 3}, with ...''. Although this design makes the sentence less fluid, participants appreciated the groundedness:

% \quo{... it doesn't just say like oh pivot index and stuff [but] like explains to me with number.}{\xx}

% 
% 
% 

\subsubsection{Formulating questions}
Despite the usefulness of grounding via questions, 
% The challenge of QA-based explanation is not entirely about the models.
a number of participants expressed difficulty in coming up with what questions to ask in the first place or formulating what they wanted to ask into a question.
\inquo{I had to think about, `How do I ask that [question]'.}{1}
\inquo{I don't really know how to word my questions like I asked it and then once it answers if I'm still confused, maybe you can like ask further questions.}{5}
Participants suggested that XAlgo could guide a user to ask questions, or how to ask questions more strategically, \eg \inquo{advise people to ask for definitions first}{1}.

{\bf Design implications}: Similar to the recent work \cite{liao2020questioning} by Liao \etal, we should develop a database of questions so that the system can suggest questions to a user either as a `warm-up' exercise, or to assist them when they struggle to formulate their own question.
With a sufficient amount of user data, a system can employ a recommender that suggest questions based on the current state of the algorithm and what have been asked by the user.

\subsubsection{Vocabulary}
At the beginning of each session, vocabulary was a main obstacle for participants to fully comprehend XAlgo's answers. 
\inquo{... they confuse me just because I think, the vocab[ulary] I didn't totally understand yet; but now that I do understand the vocab it's a lot easier to understand.
}{1}
% 
% \quo{... a lot of the answers it gave were really, um, good answers but before learning like all of the terms in the pseudocode, I didn't totally understand answers at the time}{\xx}
% 
In this study, XAlgo generated answers from the algorithm's pseudocode and comments provided by VisuAlgo, and used the majority of the terms as they are, \eg subarray, partition, and pivot. We did prepare a concept table in case participants asked about these algorithmic terms; yet we observed few participants would ask additional questions just to clarify terms used in the previous answer.

{\bf Design implications}: To help users overcome the initial gulf of vocabulary, future design can hyperlink algorithmic terms to their definitions, or more proactively explain these terms when the user first encounters them, or, as P1 pointed out, even encourage participants to ask about definitions first.

\subsubsection{Information processing} 
Participants pointed out that XAlgo's answer was \inquo{too wordy}{7} and \inquo{gives details but too much details}{2}. 
This problem stemmed from a design choice---we traded off conciseness with completeness to ensure that the answer includes all the necessary details.

{\bf Design implications}: Often there are multiple ways of answering one question (\eg \fgref{scenario}). To mitigate users' information processing load, some participants suggested a better structure, \eg \inquo{break it up into more sentences}{4}.
% , \inquo{broken down into like more basic parts}{\xx}.
Perhaps another solution---and a greater challenge---is to deliver a lengthy answer conversationally, allowing users to ask follow-up questions that gradually reveal `the complete picture'.

\subsubsection{Directness} 
Related to information processing, participants tended to favor direct answers whenever possible, \eg a yes/no to a Confirmation question.
\inquo{... gave me like clear-cut answers and gave me like a yes and no one. And I wanted a yes or no ...
}{7}
For the other types of answers, participants felt that useful and necessary information was provided but not in a direct manner, \eg \inquo{in a roundabout way}{6}:
\inquo{... it does like give you useful answers even though if they're not like the ones that directly answered the question you asked.}{6}

{\bf Design implications}: Indeed, XAlgo's answer generation focuses on retrieving question-relevant information. However, there still remains a gap between having the requisite information as-is and delivering such information in a way that is directed to answering the user's question. Future work can approach this challenge by learning from empirical examples (similar to \cite{Ehsan2019}) or using data-driven generative methods \cite{dai2017towards}.

\subsubsection{Naturalness} 
The naturalness of answers was mentioned a few times amongst participants: while they understood the utility of  the answer to provide useful and accurate information, the unnaturalness was also fairly noticeable (as shown in the wide distribution of ratings in \fgref{fig_xalgo_feedback}):
\inquo{... it answers everything accurately and it gives the information that I asked for but it does so like sounding more like a glossary like a dictionary ...}{8}
\inquo{... just use the vocabulary that you need anyway, so it's not going to like, you know, sound like a conversation ...}{4}
\inquo{... like a robot's answers ... If I asked someone to explain it, it wouldn't give me all this.}{2}

{\bf Design implications}: 
% The pursuit of a highly human-like natural language response has been a long-standing challenge that is yet to be fully realized in the future.
In their paper on mixed-initiative reasoning \cite{tecuci2007seven}, Tecuci \etal argued that ``human-agent communication needs to be as natural and efficient as possible for the human, and as complete and unambiguous as possible for the agent''.
%  and has been shown as a suboptimal approach. 
% As a compromise, people do develop `unnatural' ways of `speaking' to a system as they become familiar expert users. 
Rather than pursuing a fully human-like language, a pragmatic solution is for QA-based XAI systems to define a formal language that can jointly achieve the goal of explanation while being learnable by a human user.
% , even if it means being unable to sound like a human does.

\subsection{How well participants achieved the learning tasks}
Both XAlgo and the control group participants made progress but still struggled in understanding the algorithm.

\subsubsection{Quizzes results}
One experimenter graded participants' quizzes based on pre-generated keys. We conducted Mann-Whitney U tests, the results of which show that there is no significant difference between XAlgo and the control group in either multiple-choice scores (U = 38.0, p = 0.863) or sorting walkthrough scores (U = 40.0, p = 1.000). The overall average combining the two groups is 4.56/6 for multiple choice and 5.83/10 for sorting walkthrough. 

{\bf Design implications}: In hindsight, the quiz questions overall might have been too difficult for a user who had just started to learn an advanced sorting algorithm less than 30 minutes before. However, the more fundamental challenge, we believe, is that to achieve an advanced level of understanding an algorithm, a user would need {\it active} guidance. XAlgo's approach, on the other hand, remains passive. As discussed in the previous subsection, oft-times participants did not know what questions to ask that would advance their understanding of an algorithm. Perhaps a future design of QA-based XAI systems (at least in a learning scenario) should be able to proactively prompt and guide users to ask meaningful questions, or even generate questions to quiz users as a way to provide active guidance.

% \xac{}{more discussion}

% When asked whether they agreed that the learning task was easy (1: strongly disagree; 7: strongly agree), XAlgo users rated an average of 4.11 ($\pm$1.17) and the control group 3.11 ($\pm$1.05)---there was no significant difference (U = 21.00, p = 0.094). When asked about \xx, XAlgo users rated 5.4 ($\pm$1.88), which is significantly higher (U = 24.0, p = 0.161) than the control group (4.4$\pm$1.33). XAlgo users also considered the learning experience more enjoyable than the control group, rated 5.9$\pm$1.17 and 4.7$\pm$2.18, respectively ().

% shows that XAlgo users scored significantly higher than VisuAlgo users in both the multiple-choice problems (<p values> etc.) and the sorting walk-through problem (<p values> etc.).
% <interesting stats about how users interact with xalgo>

\subsubsection{Learning experience: understanding, difficulty \& enjoyment}
We asked participants from both XAlgo and the control groups to rate and comment on their overall learning experience via interacting with the system. \fgref{fig_learning_experience} summarizes the quantitative results. Overall, XAlgo's ratings on easiness, understanding and enjoyment were at least one point higher; the differences, however, were not statistically significant.

% \fgref{fig_learning_experience} shows participants' rating on the easiness, undertandability and enjoyment of their learning process.

% \fg{fig_learning_experience}{fig_learning_experience}{0.6}{Participants' response to the overall learning experience: in all three questions, XAlgo scored at least one point higher; the differences, however, were not statistically significant.
% }

To further obtain qualitative insight, we conducted a brief semi-structured interview asking each participant to elaborate their rating on the three questions.
%  and summarize the findings below. 
% We transcribed participants' responses as well as their thinking-aloud while interacting with the system. 
We made two observations that illustrated the difference of participants' experience of XAlgo compared ot the control condition. 
% We perform a word frequency analysis\footnote{We removed common words as well as words unrelated to understanding algorithms} to help us detect whether there is any meaningful differences between the two groups.

\one We noticed that only the control group participants expressed `surprise':
\inquo{I thought I understood the algorithm, but I don't think I did fully.}{12}
\inquo{... you think you got it and then it's actually not right}{13}
XAlgo's ability to answer participants' questions allowed them to verify issues that they were unsure about, whereas in the control group participants had no way of asking questions and had to make their own assumptions, at times causing `surprises' when they found out that their assumptions were actually wrong.

\two The control group participants wished there were explanations of the VisuAlgo animation.
\inquo{[I would have liked] more explanation on the code---I dont know what's what}{10}
\inquo{[If] I had more guidance alongside with them and I think [I] would be able to figure it out. And probably more time. But someone like explaining it to me would be better.}{12}

% Only XAlgo participants mentioned wishing to have more time: 

% \quo{Oh, yeah, this was fun except for the time constraint and which was stressing me out, honestly.}{\xx}

% \quo{... I feel like it was a little more time than this would be like really easy.}{\xx}

% This indicates a trade-off of XAlgo's QA-based explanation approach, that users need to spend extra time formulating and typing questions, as well as interpreting answers.

% time

% XAlgo participants are marked as PX and the control group as PC.

% {\bf Understanding} Participants from both groups felt they understood the algorithm better, although with limited confidence, over the course of interaction. XAlgo is one point higher in the average rating, although the difference is not statistically significant, which is consistent with participants’ accounts:

% \quo{... don't think I could confidently say I understand it but I think like better than where I began with.}{X\xx}

% \quo{I understand it, but there's still like minor nuances that I don't really understand.} {X\xx}

% \quo{I feel like I understand it but like not perfectly, but I definitely understand it more than I like when I came in.}{C\xx}

% \quo{I understand it better than I did when I first looked at it, but I would not say I know the algorithm.}{C\xx}

% To 
% {\bf Difficulty}

% {\bf Enjoyment}
% what's the same

% wish explanation

\section{Limitations, Discussions and Future Work}
% As XAlgo is part of effort of advancing XAI, we discuss design recommendations based on our research thus far.

\subsection{Practicality of Implementing XAlgo}
% We took a Wizard-of-Oz approach to validate how well the formal model of XAlgo works should it be implemented. 
For future work to actually implement XAlgo, one major task is to realize the requisite NLP capabilities. Specifically, question type classification can be implemented based on classification techniques such as SVM or logistic regression;
% \cite{berant2013semantic, andreas2016learning}
answer generation can be implemented based on either hand-crafted templates specific to each explanation type or by taking a data-driven generative approach (similar to \cite{Ehsan2019}). 
Perhaps the main challenge is generating conceptual or state-related annotations. There are two solutions for future work: 
    \one promoting and enabling a best practice for developers to create such annotations as they author an algorithm, which can be supported via a tool that extracts structures from static analysis of the code and prompts developers to enter comments for specific states;
    \two for algorithmic components that are commonly used (\eg sorting routines), future work can take a data-driven approach to mine large corpora of publicly available source code (\eg \cite{wong2015clocom}) and aggregate developers' comments as material for answer generation.
    % \item 
% \end{itemize}

\subsection{Extending the XAlgo Approach for Data-Driven Statistical Algorithms}
Although data-driven, statistical algorithms (\eg neural networks) can also be thought of as a network of states, the fundamental difference is that such states (\eg neurons in a neural network) are too low-level to represent high-level logic. Thus it would be futile to naïvely apply XAlgo to locate, \eg the excitation of a specific neuron. Rather, we expect XAlgo to be combined with existing research on high-level visualization of a neural network, \eg Olah \etal's approach that can indicate patterns of object recognition at various levels (\eg neuron, channel, layer) \cite{Olah2017}. In particular, XAlgo can consider these visualization building blocks the same way it treats different states in an HDAG and provide an interactive mechanism for users to navigate such visualizations.
% , \eg deep neural networks. Even though this is a different family of algorithms, XAlgo's question classification step still applies. The main challenge is getting the answer. XAlgo's state-based method would require a mechanism to map a user's question to a part of the network, either a specific layer or a specific neurons. Once the specific elements of the network is localized, methods \eg Olah's neuron visualiation \cite{Olah2017} might be used to generate answers to the user's question.

% We use this section to address our research questions using our findings and discuss how XAlgo can function in other use cases, as well as how different aspects of our system can be improved.
\subsection{An XAlgo Approach to Provide Interactive, User-Centered XAI}
XAlgo represents an underexplored XAI approach where the explainable medium is no longer static representation of an algorithm, but an interactive dialog that allows users to ask for specific kinds of explanations they deem useful. In this way, a system can provide a user-centered explanation, rather than explanations derivative of an algorithm's existing representations. Specifically, XAlgo works best for explaining a specific class of AI-enabled systems that exhibit distinctive state changes, \eg in a navigation scenario,
% where states are constantly changing as a response to environmental or user input. For example, consider GPS navigation. Using XAlgo, 
a driver can ask ``Why not taking the highway?'', which is classified as a Contrast question.  Through the navigation algorithm's internal logic, the system identifies that choosing local streets vs. highway is to achieve an immediate goal of avoiding traffic, thus answers ``If taking the highway, you will not avoid traffic.'' 
% The user follows up by asking ``Is there a lot of traffic on the highway?'' XAlgo identifies the question as Yes-No, checks the \insec{objects} in the navigation algorithm related to traffic, and answers ``No. Traffic is normal.'' The user realizes that the route was probably based on outdated traffic information and decides to take the next highway entrance. It turns out the traffic is indeed back to normal.

\subsection{Exploring More Algorithms and More Scenarios}
% XAlgo can be used for deterministic algorithms that behave like a finite state machine. 
We focus on one specific example---QuickSort---for illustrating the mechanism of XAlgo and for enabling a specific application scenario. One promising future work is to develop XAlgo into a toolkit that developer can use for explaining their own programs, which also allows for testing XAlgo's model on a larger variety of algorithms.
% XAlgo is generalizable and application-agnostic. 
While future work should validate the generalizability of XAlgo's approach in wider range of scenarios, in this paper we chose to focus on one specific application---learning a sorting algorithm---so that we can perform an in-depth study to gather empirical evidence on how well XAlgo works: specifically, what kinds of questions people asked, whether XAlgo provides value-added responses, and what remain as challenges to bridge users' gulf of understanding algorithms.

\begin{acks}
This work is funded by the National Science Foundation under grant IIS-1850183 and the Hellman Fellowship.
\end{acks}

\bibliographystyle{ACM-Reference-Format}
\bibliography{references,mypubs}

% \appendix

% \section{Supplementary Materials}

\end{document}